\documentclass[12pt]{revtex4}

\usepackage{epsfig,latexsym,graphicx,amsfonts,bm}
\newcommand{\beq}{ \begin{equation} }
\newcommand{\eeq}{ \end{equation} }
\newcommand{\beqs}{ \begin{eqnarray} }
\newcommand{\eeqs}{ \end{eqnarray} }

\newcommand{\f}[2]{\frac{#1}{#2}}

\newcommand{\Dp}[2]{\ensuremath{\f{\partial#1}{\partial#2}}}
\newcommand{\Dpt}[1]{\ensuremath{\Dp{#1}{t}}}
\newcommand{\DDp}[2]{\ensuremath{\f{\partial^2#1}{\partial{#2}^2}}}

\begin{document}

\title{Surface Waves on a Semi-toroidal Water Ring}

\author{Sunghwan Jung $^{1}$}
\author{Erica Kim $^{1}$}
\author{Michael J. Shelley $^{1}$}
\author{Jun Zhang $^{2,1}$}
\affiliation{$^{1}$ Applied Mathematics Laboratory, Courant
Institute of Mathematical Sciences, New York University, 251
Mercer Street, New York, New York 10012, USA \\
$^{2}$ Department of Physics, New York University,
4 Washington Place, New York, New York 10003, USA}%


\begin{abstract}
We study the dynamics of surface waves on a semi-toroidal ring of
water that is excited by vertical vibration. We create this
specific fluid volume by patterning a glass plate with a
hydrophobic coating, which confines the fluid to a precise
geometric region. To excite the system, the supporting plate is
vibrated up and down, thus accelerating and decelerating the fluid
ring along its toroidal axis. When the driving acceleration is
sufficiently high, the surface develops a standing wave, and at
yet larger accelerations, a traveling wave emerges. We also explore
frequency dependencies and other geometric shapes of confinement.
\end{abstract}

\maketitle

In 1831, Faraday first observed that surface waves on a vibrated
fluid volume oscillate at half the driving frequency
\cite{Faraday31}. These {\it Faraday waves} demonstrate the
surface instability caused by strong oscillatory accelerations
exerted on the fluid volume. Benjamin \& Ursell showed that the
amplitude of a surface eigenmode obeys the Mathieu equation
\cite{Benjamin54,Miles90}. It follows that the superposition of
multiples of harmonics and subharmonics are also solutions to the
Mathieu equation. In the presence of finite viscosity, however,
the subharmonic response is dominant
\cite{Kumar96,Cerda98,Chen99}.

Previous experiments have revealed that patterns of various
symmetries, such as striped \cite{Edwards93,Daudet95}, triangular
\cite{Muller93}, square \cite{Lang62,Ciliberto91,Edwards94}, and
hexagonal \cite{Kumar95,Kudrolli96}, can be excited on the free
surface of a fluid layer. Quasi-one dimensional surface waves have
also been studied in both narrow annular and channel geometries
\cite{Keolian81,Douady90,Vega01,Mancebo02}. In these experiments,
the standing waves often interact with the meniscus that forms at
the bounding walls. The contact point and length-scale of this
meniscus continually change as a result of the vertical
oscillations. Consequently, the meniscus emits waves towards the
bulk \cite{Douady90}. In other experiments, to remove this
meniscus effect, the contact point is pinned on a sharp edge or
brim \cite{Benjamin79,Martel98,Westra03}.

A curved fluid surface, such as on a hemi-spherical drop
\cite{Lamb32,Bisch82,Wilkes97,Frohn00,Noblin04}, has been used to
study surface waves in a confined geometry without bounding walls.
Vibrating the drop causes waves to form on the curved surface, and
for high external forcing, droplets are ejected. This process,
termed {\it atomization}, provides one way to create a spray
\cite{James03a,James03b}. The instability of the drop's
semi-spherical surface has been utilized for spray cooling,
mixing, and humidification.

Faraday instability has been well studied both in flat fluid
surfaces \cite{Edwards94,Daudet95,Kumar95,Kudrolli96} confined by
the lateral boundaries and in sessile (pendant) drops
\cite{Bisch82,Wilkes97,James03a,Noblin04}. The differences between
the two cases are their aspect ratios and boundary conditions
(moving lateral contact lines or the capillary pinning). The
geometry of our interest lies between the two cases: the
semi-toroidal geometry is spatially extended with high aspect
ratios (system size/capillary length) and pinning boundary
conditions apply. It is however much less studied compared to the above
two settings. Here, we employ a hydrophobic/hydrophilic patterning
of a surface to confine a water volume to a specific region. In
particular, if part of a hydrophilic substrate (low surface
energy) is covered by a hydrophobic coating (high surface energy;
we use Fluorothane ME; Cytonix Corp.), water is prevented from
spreading beyond the hydrophilic region.  The boundary between the
two regions acts as an edge upon which the fluid contact point is
pinned.

In the experiment we report here, the hydrophilic region is a thin
annulus, and water placed there forms into a semi-toroidal volume
with two pinned contact lines.  We find that surface waves first
develop along the centerline of the torus, which allows us to
study quasi-one dimensional waves on a curved surface.

\begin{figure} 
    \centering
        \includegraphics[width=.5\textwidth]{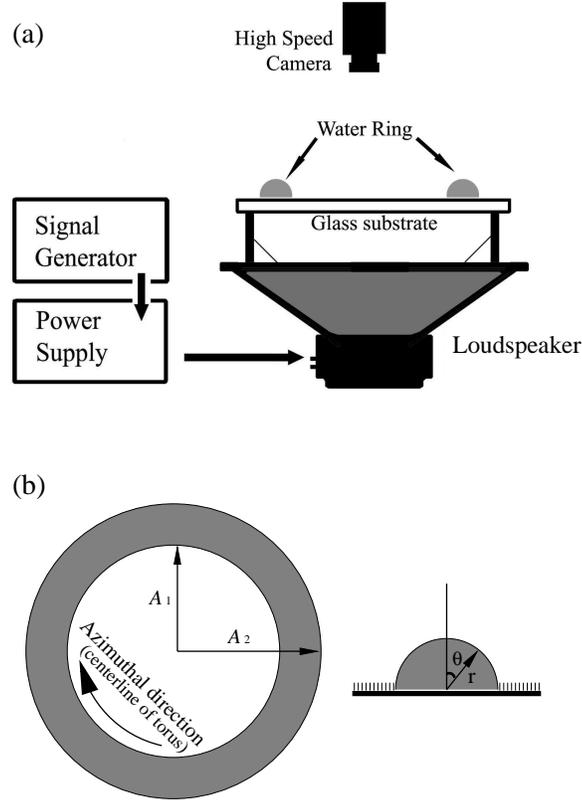}
\caption{(a) Side-view of the experimental set-up: a semi-toroidal
water-ring on the hydrophobic patterned glass plate that is
vibrated vertically by a loudspeaker. (b) Top and cross-sectional
views of the water-ring. }
    \label{fig:apparatus}
\end{figure}

The schematic of our experimental setup is shown in
Fig.~\ref{fig:apparatus}.  The annular hydrophilic region has
inner radius $a_1 = 2.5$ cm and outer radius $a_2 = 3.5$ cm. The
patterned glass plate is rigidly connected to a speaker, which
oscillates the plate vertically at a controlled driving frequency
and amplitude.  A high-speed video camera is centered above the
plate along the axis of the annulus. Since the semi-torus of water
has a cross-sectional radius ($r\approx 0.5$ cm) that is
relatively small in comparison to the distance between the plate
and the camera, bright regions of the water ring seen in
top-viewed video frames correspond to local extrema of the fluid
surface height.

\begin{figure} 
    \centering
    \includegraphics[width = .4\textwidth]{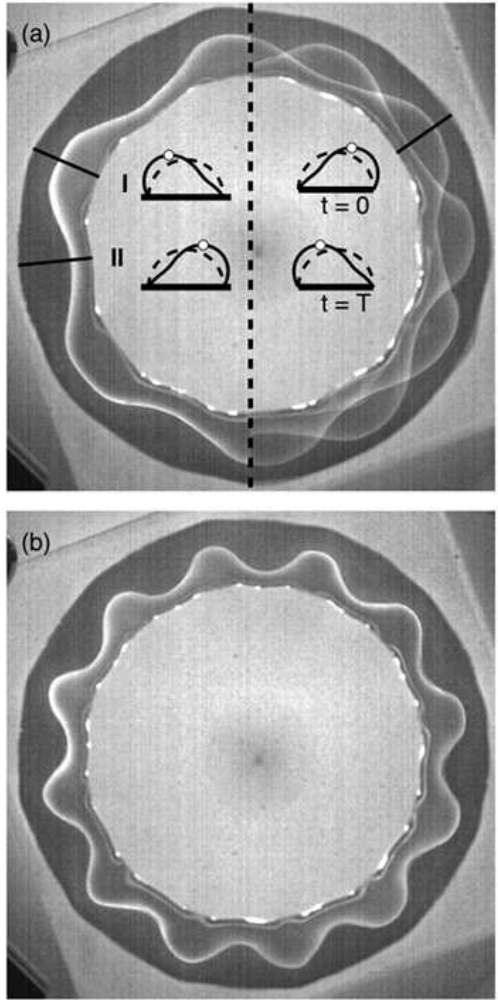}
\caption{Images of surface waves emerging from the water ring. At
the driving frequency 22 Hz, 12 nodes are excited in the azimuthal
direction (a), whereas 24 nodes appear at 36 Hz (b). In (a), the
cross-sectional shapes along lines {\bf I} (referred to as the
{\it peak}) and {\bf II} ({\it trough}) are shown schematically.
The dashed curves represent the fluid surfaces in equilibrium.
White dots represent the cross-section position of the bright
curve in a top-viewed image. The right half of image (a) is an
overlap of two photographs differing by the period of excitation
$T$ (=1/22 sec).}
    \label{fig:waves}
\end{figure}

\begin{figure} 
    \centering
        \includegraphics[width=.47\textwidth]{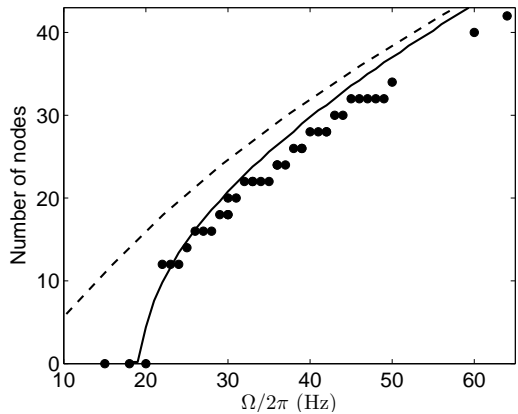}
\caption{Number of nodes around the annulus observed near the onset of
wave formation. Dots are from experiments.
The solid line is a result of the inviscid coupled linear model
and the dashed line is from the classical dispersion relation
($(\Omega/2)^2 =  g k + \sigma k^3/\rho$). }
    \label{fig:instab1}
\end{figure}

We use a systematic procedure to define the onset of surface
waves. At fixed driving frequencies $\Omega/2 \pi$ between 20 Hz
and 65 Hz, the amplitude of oscillation is slowly increased until
the fluid surface loses stability to azimuthally modulated
standing waves. Distinct wave patterns are observed at half the
driving frequencies. Figures~\ref{fig:waves}a and b show these
standing waves for two different driving frequencies. At 22 Hz
(a), the wave pattern has 12 nodes (that is, 12 crossings of the
wave peak along the annular mid-line).  The left half of
Fig.\,\ref{fig:waves}(a) shows a snapshot of the surface wave,
whereas the right half is the image of two superimposed surface
waves in different phases. In the snapshot, the cross-sectional
shapes, or onset modes, of the fluid surface are shown as insets
when cutting across the {\bf I} and {\bf II} lines. The overlapped
image demonstrates that standing waves are formed at half the
driving frequency (22 Hz) near the onset. Increasing the driving
frequency to 36 Hz increases the number of wave nodes to 24 (see
Fig.~\ref{fig:waves}b). Near the onset, as the driving frequency
increases, the number of nodes increases monotonically (and
discretely, due to the geometry), as is shown in
Fig.~\ref{fig:instab1}.

\begin{figure} 
   \centering
      \includegraphics[width=.47\textwidth]{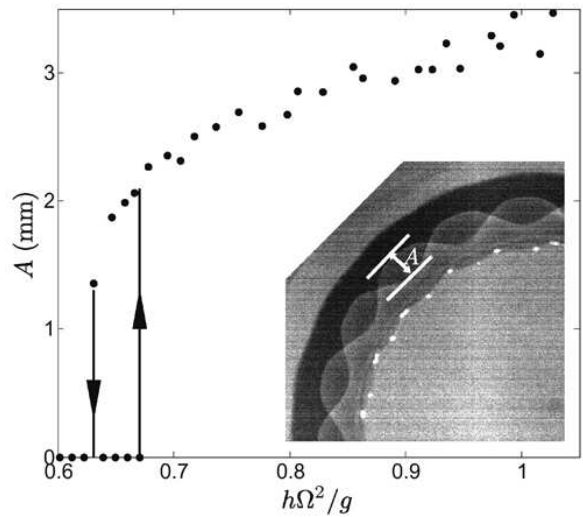} 
\caption{Lateral amplitude of oscillation vs. non-dimensional
acceleration ($h\Omega^2/g$) at $\Omega/2 \pi =$ 30 Hz. $A$ is the
transverse amplitude as indicated in the inset. This
response curve shows hysteresis over driving acceleration between
0.63 and 0.67 $g$.  }
    \label{fig:hyster}
\end{figure}

To describe the instability to the azimuthally modulated surface
waves 
seen in Fig.~\ref{fig:waves}, we consider an approximate 
model which captures the main aspects of our observations.
First, we neglect viscous effects. Second, instead of using 
the exact semi-toroidal geometry for our
calculations, we neglect the annular curvature and use a
cylindrical geometry. 
Finally, we note that the capillary length($\sqrt{2 \sigma/ \rho
g}$) in our experiment is $\sim$ 0.3 cm slightly smaller than the radius $r
\sim 0.5$ cm, and so we expect the base state (i.e. the surface profile without azimuthal 
modulation) to have a flattened, temporally oscillating surface profile. 
Solving for the temporal evolution of the surface profile, even azimuthally
unmodulated, is a nonlinear problem. Hence, for simplicity, 
we assume that the base state is close to a semi-circle with constant radius
$R$. Small fluctuations of the surface can then be expressed as $R
+ \zeta(\theta)$, where $\zeta(\theta)$ is a small radial displacement deviating
from $R$. 
This assumption leads to a simple linear analysis for the perturbed fluid surface.
A fully nonlinear, numerical treatment of the oscillations and Faraday 
instabilities of a driven viscous drop has been considered by James et al \cite{James03a}.

A linearized form of Bernoulli equation with a velocity potential
$\phi$ \cite{Benjamin54} in a moving frame is
\beqs
&g_e \zeta \cos \theta &+ \Dpt{\phi} - \f{\sigma}{\rho} \left(
\f{\zeta}{R^2}+ \f{1}{R^2} \DDp{\zeta}{\theta} + \DDp{\zeta}{z}
\right) = 0
\,, \label{eq:basic_eq} \\
&\nabla^2 \phi = 0 & \label{eq:sup_29} \,.
\eeqs
Here, $g_e(t) = g + a \cos \Omega t$ is the oscillating body force
where $g$ and $a$ are the acceleration due to a gravity and the
peak acceleration of the vibrating plate, respectively. By using $
v_r|_{r=R} \equiv
\partial_r \phi|_{r=R} =
\partial_t \zeta$ which neglects the nonlinear terms in the
kinematic boundary condition, we can assume solutions such as
\beqs
\zeta(\theta,z;t) &=& \sum \zeta_{mk} (t) e^{i(m \theta + k z)}
\label{eq:zeta}\\
\phi(r,\theta,z;t) &=& \sum \f{ d \zeta_{mk}(t) }{dt} \, \f{I_{m}
(k r) }{k I_m'(k r_0) } e^{i(m \theta + k z)} \,. \label{eq:phi}
\eeqs
where $k$ is the axial wavenumber ($= 2 \pi/L$; $L$ is the
wavelength) and $m$ is the azimuthal wavenumber (in the direction
of $\theta$). Here, we assume that $a/g$ is small and neglect
higher azimuthal modes ($|m|>2$) since we observed that $m=2$ is
the dominate mode near the onset of surface excitation. Using the
symmetry $\zeta_{m,k} = -\zeta_{-m,k}$, setting the unphysical
volume change term $\zeta_{0k}$ to be zero, and plugging
Eqs.\,(\ref{eq:zeta},\ref{eq:phi}) into Eq.\,(\ref{eq:basic_eq})
yields
\beq
\left( \begin{array}{cc} Z_{1k}\partial_{tt} + \sigma_{1k} & g_e(t) \\
g_e(t) & Z_{2k}\partial_{tt} + \sigma_{2k} \end{array} \right)
\left( \begin{array}{c} \zeta_{1k} \\ \zeta_{2k} \end{array} \right) =0 \,,
\eeq
where $Z_{mk}= I_{m}(kR)/kI'_m(kR)$ and $\sigma_{mk} = (\sigma/\rho)
\cdot(k^2 + m^2/R^2 - 1/R^2)$. The surface
deformation is assumed to be of the Floquet form ($ \zeta_{mk} =
e^{\tau t} \sum \zeta^{(m)}_n e^{i n \Omega t}$) where $\tau = s + i
\alpha \Omega$ \cite{Kumar96}. For subharmonic solutions ($\alpha
= 1/2$), one gets the recursion relation:
\beq
H_n \zeta^{(m)}_n = \f{a}{g}
\left(\zeta^{(m)}_{n-1} + \zeta^{(m)}_{n+1} \right) + {\cal O}
\left( ( a/g )^2 \right) \,,
\label{eq:recur}
\eeq
where
\beq
H_n = \left[ - 1 + 4 (Z_{1k} \omega_n^2 - \sigma_{1k})
(Z_{2k} \omega_n^2 - \sigma_{2k})/g^2 \right] \,
\eeq
and $\omega_n = (1/2 + n) \Omega$. The problem of stability of the
cylindrical fluid surface is reduced to finding the eigenvalues of
Eq.\,(\ref{eq:recur}) up to the order of $a/g$.

\begin{figure} 
   \centering
      \includegraphics[width=.5\textwidth]{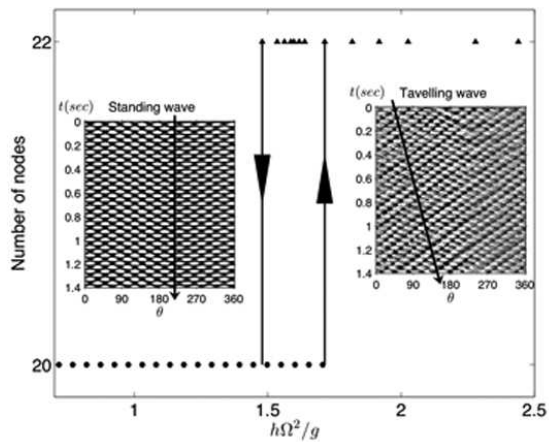}
      \caption{Number of nodes vs. non-dimensional acceleration at
$\Omega/2 \pi = 30$ Hz. As the acceleration is slowly increased,
the number of nodes changes from 20 to 22 when the acceleration is
about 1.71\,$g$. When the acceleration decreases, the number of
nodes returns back to 20 at 1.48\,$g$. The states shown in
triangles (top branch) are associated with the traveling waves.}
   \label{fig:hyster_node}
\end{figure}

\begin{figure} 
    \centering
        \includegraphics[width=.5\textwidth]{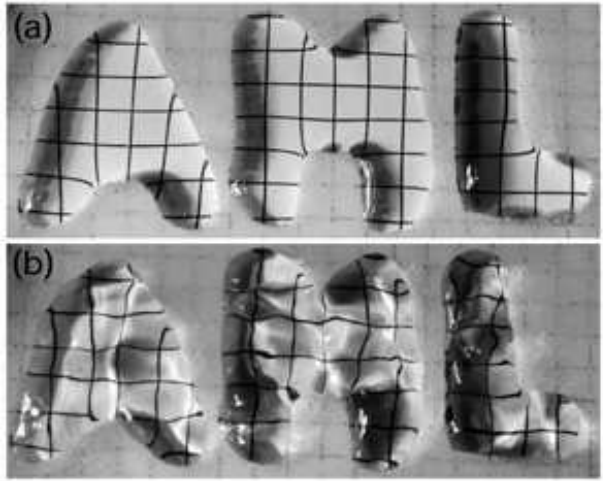}
\caption{ (a) Below threshold, stable fluid surface constrained to
three regions which are bordered by a hydrophobic coating. (b)
Above the onset, unstable fluid surface vibrating at half of the
driving frequency. At 50 Hz, fluid surfaces constrained to the
shapes ``A'',``M'',``L'' become unstable at accelerations 2.51,
2.23 and 2.64 $g$, respectively.}
   \label{fig:aml_shape}
\end{figure}

Figure\,\ref{fig:instab1} compares the observed patterns with the
inviscid coupled linear model and the classical dispersion
relation ( $(\Omega/2)^2 = g k + \sigma k^3/\rho$ ). Our coupled
model is in better agreement with our experimental observation
than the classical dispersion relation.

We also find experimentally that the onset of surface waves is
hysteretic. Figure~\ref{fig:hyster} shows the maximal displacement
$A$ in the horizontal direction between standing wave peaks over
the course of an oscillation, as a function of dimensionless
driving acceleration $h \Omega^2 /g$, where $h$ is the amplitude
of plate oscillation. Crossing a critical amplitude from below
shows the sharp transition to standing wave. This bifurcation is
subcritical, and at intermediate accelerations ($0.63\,g<
h\Omega^2< 0.67\,g$) two stable states are observed.

Figure~\ref{fig:hyster_node} shows yet another hysteretic
transition in the number of surface nodes, at a higher
dimensionless acceleration. Bistable surface waves with different
numbers of nodes co-exist between 1.48\,$g$ and 1.71\,$g$. We find
that the surface waves above this transition still oscillate at
half the driving frequency, but now have more nodes and a finite
drift speed (triangles). The inset images show the patterns of
peaks (bright) and troughs (dark) in space-time. The arrows
indicate the direction of azimuthal drift and its speed. If the
acceleration is further increased to around 4\,$g$, the traveling
surface waves become more spatially localized, and droplets are
ejected.

Our masking technique can be used effectively to make complicated
hydrophilic regions, and to investigate the dynamics of fluid
volumes trapped within them.  Figure~\ref{fig:aml_shape} shows
three separate fluid volumes bound within domains (the letter
shapes A, M and L) of different shapes at both stable and unstable
states. At fixed volume to base-area ratio of 0.16 cm$^3$/cm$^2$,
which gives a fixed height for all shapes slightly above 0.18 cm,
instability occurs at different thresholds for the different base
shapes. The shape that with high aspect ratio (e.g. letter ``L'')
appears to have the highest threshold to instability. The top
image shows these volumes at rest, where the transparent glass
surface of the hydrophilic region reveals a square grid beneath.
The bottom image shows the surface wave dynamics of these fluid
volumes as revealed, at least in part, by the distortions of the
underlying grid lines.

In this paper, we have studied the Faraday instability of a
semi-toroidal water volume whose surface is initially flat in the
azimuthal direction but tightly confined in the radial direction.
To constrain the fluid into a tight geometric area (an annulus),
the base plate is patterned with hydrophobic and hydrophilic
regions. This particular geometry allows us to study the surface
waves arising from a quasi-one-dimensional system with a periodic
boundary condition. Systems with high aspect ratios (system
size/capillary length), as used in our experiment, have not been
well studied previously. Despite of the high aspect ratio, we find
similar dynamics (such as the hysteretic excitation of surface
waves) as found in systems with flat surfaces. The classical
theory is modified to describe the observed dispersion relation in
our system. The result agrees with our experimental observations.
For large forcings, we observe another transition, again
hysteretic, to traveling waves with an increased number of nodes.
Such a transition has yet to be understood theoretically and is a
subject of future investigation.

Authors thank A. Libchaber, D. Hu, and V. Rom-Kedar for helpful
discussions. This work is supported by DOE Grant
DE-FG02-88ER25053.

\end{document}